\documentclass[useAMS,usenatbib,usegraphicx]{mn2e}

\newcommand{\epcs}{{\rm erg\,cm^{-2}\,s^{-1}}}

\newcommand{\nat}{Nature}
\newcommand{\apj}{ApJ}
\newcommand{\apjl}{ApJL}
\newcommand{\apjs}{ApJS}
\newcommand{\aap}{A\&A}
\newcommand{\ssr}{SSR}
\newcommand{\mnras}{MNRAS}
\newcommand{\procspie}{Proc. SPIE}
\newcommand{\pasj}{PASJ}

\newcommand{\gtrsim}{\mathrel{\hbox{\rlap{\hbox{\lower4pt\hbox{$\sim$}}}\hbox{$>$}}}}
\newcommand{\lesssim}{\mathrel{\hbox{\rlap{\hbox{\lower4pt\hbox{$\sim$}}}\hbox{$<$}}}}

\newcommand{\nburst}{246}

\title[]{Biases for neutron-star mass, radius and distance measurements from
Eddington-limited X-ray bursts}
\author[D. K. Galloway et al.]{D. K. Galloway$^{1,2}$\thanks{E-mail:
Duncan.Galloway@sci.monash.edu.au}
F. \"Ozel$^{3,4}$
and
D. Psaltis$^{3,4}$ \\
$^{1}$School of Physics \& School of Mathematical Sciences, Monash
  University , VIC 3800, Australia\\
$^{2}$Monash Fellow\\
$^{3}$Department of Physics, University of Arizona, 1118 E. 4th Street, Tucson,
AZ, USA \\
$^{4}$also Astronomy Department, University of Arizona}
\begin{document}

\date{\today}

\pagerange{\pageref{firstpage}--\pageref{lastpage}} \pubyear{2006}

\maketitle

\label{firstpage}

\begin{abstract}
Eddington-limited X-ray bursts from neutron stars can be used in
conjunction with other spectroscopic observations to measure neutron
star masses, radii, and distances. In order to quantify some of the
uncertainties in the determination of the Eddington limit,
we analysed a large sample of photospheric radius-expansion thermonuclear
bursts 
observed with the {\it Rossi X-ray Timing Explorer}. 
We identified the instant at which the expanded photosphere ``touches
down'' back onto the surface of the neutron star 
and compared the corresponding touchdown flux to the peak flux of
each burst. We found that for the majority of sources, the ratio of
these fluxes is smaller than $\simeq 1.6$, which is the maximum value
expected from the changing gravitational redshift during the radius
expansion episodes (for a $2M_\odot$ neutron star).  The only sources for
which this ratio is larger than $\simeq 1.6$ are high inclination sources
that include dippers
and Cyg~X-2. 
We discuss two possible geometric interpretations of
this effect and show that the inferred masses and radii of neutron
stars are not affected by this bias. On the other hand, systematic
uncertainties as large as $\sim 50\%$ may be introduced to the
distance determination.
\end{abstract}


\section{Introduction}

Photospheric radius-expansion (PRE) thermonuclear bursts are a key
observational tool for studies of low-mass X-ray binaries. These systems
consist of a neutron star accreting 
from a low-mass
stellar companion. Thermonuclear (type I) bursts arise when a critical
temperature and density is reached in the accumulated fuel layer, and
unstable H or He ignition takes place \cite[e.g.][]{lew93,sb03}. The
accumulated fuel then burns and is exhausted within 10--100~s.
Ongoing accretion leads to subsequent bursts separated by hours to tens of
hours (depending upon the accretion rate).

The X-ray flux early in the burst can be sufficient to exceed the local
Eddington limit at the surface of the neutron star. In that case,
radiation presure drives an expansion of the photosphere, and excess
flux is converted to kinetic and gravitational potential energy. The
expansion of the photosphere at approximately constant luminosity produces
a characteristic pattern of the X-ray spectral variation in these bursts,
giving rise to a peak in the blackbody radius at the same time as a local
minimum in the blackbody temperature. Measurement of the X-ray flux during
the PRE episode allows determination of the distance to the burst source
\cite[e.g.][]{bas84}, 
as well as the mass and radius of the neutron star \cite[e.g.][]{damen90,ozel06}
These measurements are hampered by intrinsic variation in the
Eddington luminosities between sources and even from burst to burst
(e.g. \citealt{kuul03a}; see also \citealt{bcatalog}).

One subtlety that arises for measurements of the Eddington flux $F_{\rm
Edd}$ is the precise instance during the burst at which this value is
measured. The observed flux during the PRE episode is expected to vary due
to the changing gravitational redshift resulting from the
varying height of the photosphere above the neutron star. Ideally, the
flux should be measured when the Eddington limit is first reached, just
before radius expansion commences, or alternatively when radius expansion
ceases and the expanded photosphere ``touches down'' once again onto the
neutron star. The instance when radius expansion commences is typically
difficult to identify, since it occurs during the burst
rise. Instead, the flux at touchdown, at the end of the PRE episode, has long
been thought to be the best time to measure the Eddington flux
\cite[]{damen90}.

Recently, measurements of the X-ray flux at touchdown in radius-expansion
bursts from EXO~0748$-$676 were used (along with other observational
constraints) to derive limits on the mass and radius of the neutron star
\cite[]{ozel06}. The Eddington flux 
at touchdown measured
from bursts observed by {\it EXOSAT}\/ and {\it RXTE}, $2.25\times10^{-8}\
\epcs$, is significantly lower than the maximum flux reported from the
bursts observed by {\it RXTE}, of $5.2\times10^{-8}\ \epcs$
\cite[]{wolff05}. This discrepancy \cite[also noted by][]{damen90} of more than a factor of 2, is far in
excess of that expected from the varying gravitational redshift during
radius expansion. If the touchdown flux does indeed correspond to the
Eddington limit, it is difficult to understand how the flux earlier in the
burst could exceed that value by such a large margin.

Motivated by this
result, we undertook a broader study of the relationship between the peak
radius-expansion burst flux and the flux at touchdown in different burst
sources. In
\S\ref{obs} we describe the observations and analysis of the bursts. In 
\S\ref{results} we present the results of our study. 
In section \S\ref{lowtouch} we identify sources with low touchdown fluxes
(compared to the peak fluxes), while in \S\ref{maxrad} we investigate the
effect of the maximum radius reached during the expansion.
We present the broader conclusions of this study in \S\ref{conc}.

\section[]{Observations and analysis}
\label{obs}

We used analyses of thermonuclear bursts observed by the {\it Rossi X-ray
Timing Explorer}\/ ({\it RXTE}) from the catalogue of \cite{bcatalog}.
This catalogue contains analyses of all bursts present in public data
since shortly after the {\it RXTE}\/ launch on 1995 December 30. The
primary instrument for burst studies is the Proportional Counter Array 
\cite[PCA;][]{xte96},  consisting of five
identical, co-aligned proportional counter units (PCUs), sensitive to
photons in the energy range 2--60~keV and with a total effective area of
$\approx6500$~cm$^2$. 

Data were analysed with {\sc LHEASOFT} version 5.3 (released 2003 November
17) or later; this release was the first to incorporate a significant
reduction of the geometric area of the PCUs for improved consistency with
calibration sources \cite[]{xtecal06}.
Data with the best spectral and temporal
resolution were used to extract time-resolved spectra covering the PCA
bandpass every 0.25~s near the peak of each burst, becoming longer in the
burst tail to preserve signal-to-noise.
Each burst spectrum was fitted using an absorbed blackbody, with
a spectrum extracted from a (typically) 16~s interval prior to the burst
as the background, following standard X-ray burst analysis procedure
\cite[e.g.][]{vpl86,kuul02a}.
To take into account gain variations over the life of the instrument, we
generated a separate response matrix for each burst using {\sc pcarsp}
version 10.1.

The bolometric flux at each timestep was calculated from the fitted
blackbody temperature $kT_{\rm bb}$ and radius $R_{\rm bb}$. Each burst
was inspected visually for evidence of radius expansion, specifically a
local maximum in $R_{\rm bb}$ near the time the maximum burst flux
was reached, synchronous with a minimum in $kT_{\rm bb}$. By this
criteria, \nburst\ bursts in the catalogue exhibited unambiguous evidence of
radius expansion.

We subsequently analysed the bolometric flux evolution of each of these
bursts to determine the touchdown flux $F_t$. Immediately prior to
touchdown, the photosphere is expected to be contracting back towards the
neutron star, reflected by the decreasing blackbody radius. 
The long ($\sim1$~s) observed timescale of the radius contraction
(compared to the $\sim1$~ms free-fall timescale) suggests that the
atmosphere is still primarily radiation-pressure supported during the
contraction phase.
Because of this condition, the flux remains equal to
the local Eddington limit, and the blackbody temperature must be {\it
increasing}. Immediately after touchdown, the photospheric radius is
roughly constant, but now the burning has more or less ceased and the
neutron star gradually cools, giving a decreasing flux and blackbody
temperature. We thus identify the time at which the photosphere touches
down as the time of a local maximum in the blackbody temperature,
following the time of maximum radius.

\begin{figure}
 \includegraphics[width=3.3in]{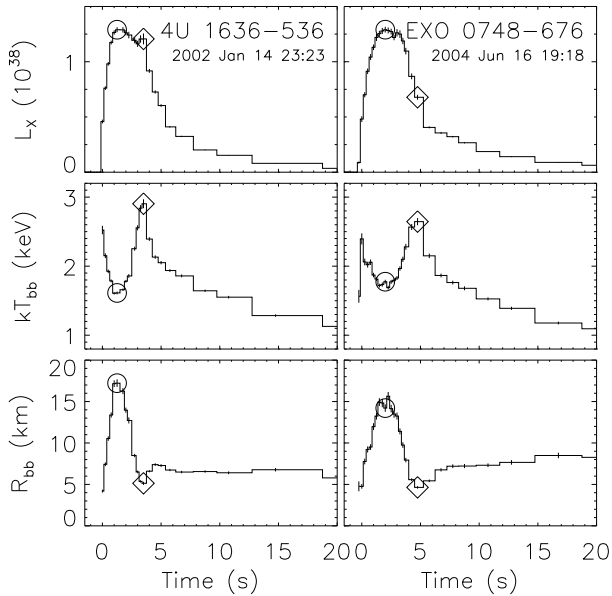}
 \caption{Example photospheric radius-expansion bursts from 4U~1636$-$536
({\it left panels}) and EXO~0748$-$676 ({\it right panels}). 
From top to bottom, the panels in each column show the bolometric
luminosity (units of $10^{38}$~erg~s$^{-1}$); blackbody temperature (keV);
and blackbody radius (km). The luminosity and blackbody radii are
calculated for a distance to 4U~1636$-$536 of 6~kpc \citep{gal06a}, and
to EXO~0748$-$676 of 7.4~kpc \citep{bcatalog}. The open circles indicate
the time of peak burst flux, while the open diamonds indicate the time
when the photosphere is assumed to have just returned to the neutron star
surface (the ``touchdown'' point). Note that the flux at touchdown $F_t$ for
4U~1636$-$536 is essentially identical to the maximum flux, while for
EXO~0748$-$676 it is approximately half that value.
  \label{bursts} }
\end{figure}

This definition is illustrated in the two examples of Fig.
\ref{bursts}. The time of touchdown can be unambiguously determined from
the time-history of $kT_{\rm bb}$; in both cases a clear local maximum can
be seen ({\it middle panels}), following the radius maximum ({\it bottom
panels}). While not all the cases were as clear-cut as these examples, we
nevertheless were able to determine the time of touchdown for each of the
radius-expansion bursts in the sample.

\section[]{Results}
\label{results}

We measured the flux at touchdown $F_t$ for each of \nburst\ photospheric
radius-expansion bursts in our sample, and calculated its ratio to the
peak flux $f=F_p/F_t$. 
For 61 bursts (around 1/4 of the sample) the maximum flux was achieved at
touchdown, so that $F_p=F_t$ and $f=1$. More generally, $F_t$ was
consistent with $F_p$ to within the $3\sigma$ confidence limit (based on
the uncertainties of $F_p$ and $F_t$) for 184 of the bursts, or almost 3/4
of our sample (the burst shown in the left panels of Fig. \ref{bursts} is
such an example). In many cases this commensurability of the peak and
touchdown fluxes arises from a common behaviour in PRE bursts, where the
flux continues to increase following the time the maximum radius is
reached \cite[as noted for 4U~1636$-$536 by][]{seh84}.
We note that $f>1$ for all three radius-expansion bursts from
EXO~0748$-$676, as we expect based on the comparison of the published
fluxes by \cite{ozel06} and \cite{wolff05}. Thus, we made a closer
examination of what conditions give rise to values of $f>1$.

\subsection[]{Sources with low touchdown fluxes}
\label{lowtouch}

Radius-expansion bursts from eighteen different sources gave rise to
values of $f>1$. In many cases these values were only a few percent in
excess of 1; by comparison, the $f$-value for two of the three bursts from
EXO~0748$-$676 were in excess of $f>1.9$, and the maximum value over all
the bursts was $f=6.72$. Bursts with $f>1.6$ arose from just seven
sources:
EXO~0748$-$676,
MXB~1659$-$298,
4U~1916$-$053,
GRS~1747$-$312,
XTE~J1710$-$281,
4U~1746$-$37 and
Cyg~X-2.
The remarkable feature of this selection on $f$ is that six of the seven
sources listed above (excluding Cyg~X-2) all exhibit regular X-ray dips
\cite[see e.g.][]{lmxb01} attributable to absorption from structures in the
accretion disk, occuring at preferential orbital phases. Since the
accretion disk is usually thought to have limited vertical extent, the
presence of dips is normally taken as an indication of high system
inclination, i.e., $i\approx90^\circ$. We thus conclude that systematically
large $f$-values overwhelmingly (but not exclusively) occur in bursts from
systems with high inclination; we discuss the remaining example, Cyg~X-2,
in 
\S\ref{conc}.

\begin{figure}
 \includegraphics[width=3.3in]{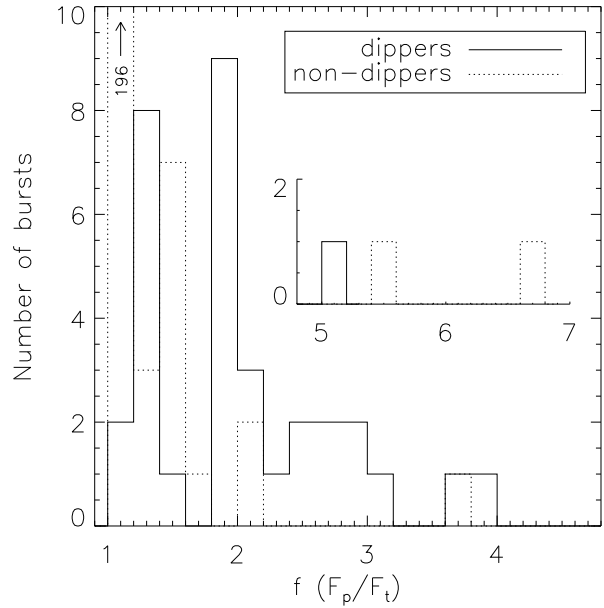}
 \caption{The distribution of $f=F_p/F_t$, the ratio of the peak flux to
the touchdown flux, for \nburst\ radius-expansion bursts in our sample.
The sample has been divided based on the source type; we plot the
distribution for the dippers
(EXO~0748$-$676,
MXB~1659$-$298,
4U~1916$-$053,
GRS~1747$-$312,
XTE~J1710$-$281 and
4U~1746$-$37, {\it solid histogram}) separately from the non-dipping
sources ({\it dotted histogram}). The overwhelming majority of bursts from
the non-dipping sources have $f<1.2$; we have indicated this in the
histogram by labelling the first bin (which is cut off) with the data
value. The inset shows the three bursts with $f>4$ at the extreme end of
the $f$ distribution.
Note that the only non-dipping source producing bursts with $f>1.6$ is
Cyg~X-2. 
  \label{fdist} }
\end{figure}

We thus divided the sample between dipping and non-dipping sources, and
calculated the combined distribution of $f$ from each sample (Fig.
\ref{fdist}). While there is substantial overlap between the two
distributions, the preponderance of bursts with $f\approx1$ for the
non-dipping sources makes the two distributions highly discrepant.
The K-S test value for the two samples is 0.886, with associated
probability $2\times10^{-21}$; this indicates unambiguously that the two
samples arise from different populations. For the dippers, the mean
$f=2.1\pm0.9$, so that the flux at touchdown was, on average, less than
half the maximum flux.

\subsection[]{The effect of the maximum radius in dipping sources}
\label{maxrad}

While the association of low touchdown fluxes with high system inclination
is clear, the underlying cause (as well as the wide range of the $f$
values for the dipping sources) begs an explanation. The large $f$-range
was most notable for MXB~1659$-$298, where the $f$-values were between 1.3
and 5.1 for the 12 radius-expansion bursts in the sample.

\begin{figure}
 \includegraphics[width=3.3in]{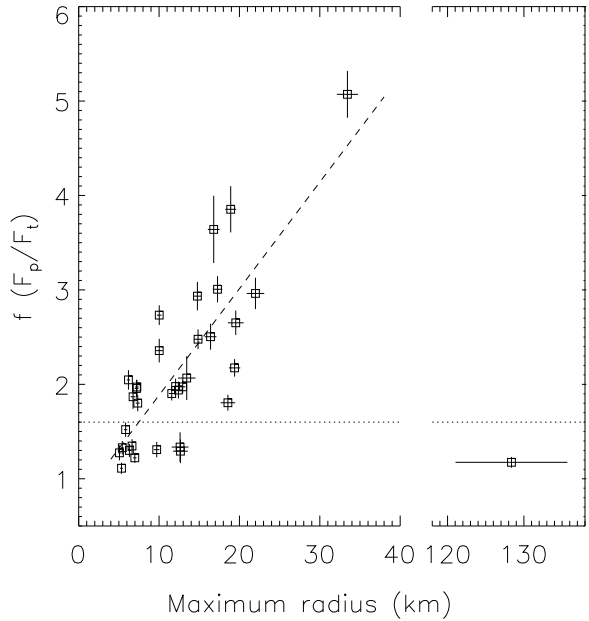}
 \caption{The ratio of peak to touchdown flux $f=F_p/F_t$ as a function of
the derived maximum photospheric expansion radius for a sample of bursts
observed from 6 dipping sources:
EXO~0748$-$676,
MXB~1659$-$298,
4U~1916$-$053,
XTE~J1710$-$281,
GRS~1747$-$312 and
4U~1746$-$37. The dashed lines shows
the best linear fit excluding the single outlier, an extremely
intense burst from GRS~1747$-$312 which reached an estimated maximum
radius of almost 130~km.
  \label{fvsr} }
\end{figure}

The high end of the distribution in the ratio $f$ is too large to be
explained by the change in the gravitational redshift during the
expansion of the photosphere. For example, for a 10~km, $2 M_\odot$
neutron star, the flux ratio arising from redshift changes can be at
most equal to 1.6. Instead, the clear separation between the
$f$-distributions of the dippers and non-dippers strongly suggests a
geometric origin of the $f$-ratios.

A geometric interpretation is further supported by the correlation
between the ratio $f$ observed in dippers and the maximum radius
attained by the photosphere during the radius expansion episode
(Figure~\ref{fvsr}).
In order to calculate the maximum radius, we
corrected for the distance to the dipping sources based (in the case of 
EXO~0748$-$676,
MXB~1659$-$298,
4U~1916$-$053 and
XTE~J1710$-$281) on the distance derived from the peak flux of the PRE
bursts \cite[]{bcatalog}. In the case of GRS~1747$-$312 and 4U~1746$-$37,
we used instead the estimated distances to the host globular clusters,
Terzan~6 (9.5~kpc) and NGC~6441 (11~kpc), respectively \cite[]{kuul03a}.
We found that, for all but one burst, $f$ is strongly correlated with the
maximum radius (Fig. \ref{fvsr}). Excluding a single outlier \cite[an unusually
intense burst from GRS~1747$-$312; see also][]{zand03a}, we find a
Spearman's rank correlation of 0.742, significant at the 4.2$\sigma$
level.

\begin{figure}
 \includegraphics[width=3.3in]{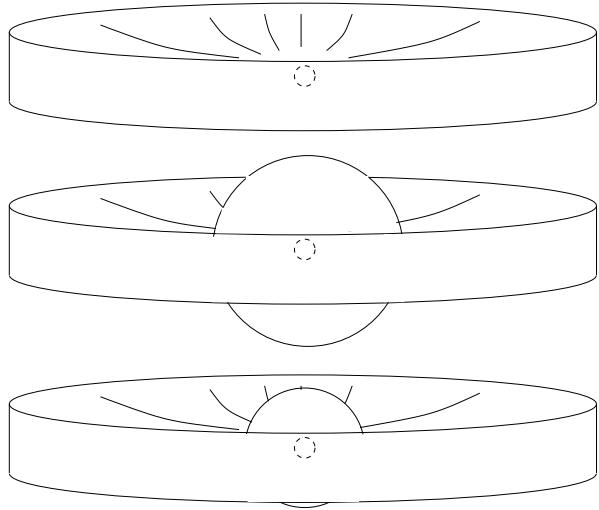}
 \caption{Schematic showing the suggested geometry giving rise to the low
touchdown flux values (not to scale). Prior to burst ignition ({\it top
panel}) the neutron star ({\it dashed small circle}) is partially obscured
by the edge of the accretion disk. When a burst ignites and causes the
photosphere to expand, the expansion out of the disk makes the photosphere
more visible. As the photosphere collapses back towards the neutron star
({\it lower panel}) obscuration by the disk becomes increasingly
important.
  \label{schem} }
\end{figure}

We suggest that the preferential occurrence of high $f$-ratios in dippers
as well as the correlation shown in Figure~\ref{fvsr} can be accounted for
either by a changing partial obscuration of the neutron star and its
expanding photosphere by the disk during the radius expansion or by a
varying reflection of the burst emission off the far side of the disk.
In the first case, which is illustrated in Figure~\ref{schem}, it is plausible
that 
even during non-dipping intervals, the neutron star in high-inclination
systems is subject to a certain degree of obscuration and hence absorption
by the disk material.
The expanded photosphere
present during the radius expansion episode increasingly protrudes above
and below (relative to our line of sight) the disk, so that the emitted
flux during radius expansion is less subject to absorption by the disk
material, and the apparent flux is commensurably larger. When the
photosphere touches down onto the neutron star again, the absorption
increases once again, giving rise to a significantly lower touchdown flux.
This requires a
particularly finely-tuned inclination, in which the viewing angle of
the observer grazes the accretion disk at its maximum thickness.

In the second interpretation, as the photosphere expands, the far side
of the disk intercepts and reflects an increasing fraction of the
emitted luminosity. This leads to an artificial increase of the
observed flux at the peak of the radius-expansion episode, while the
touchdown flux corresponds to the true Eddington limit. Albeit
plausible, the large observed $f$-ratios require, in this
interpretation, a large amount of reflection off the disk, which can
by achieved only in the presence of significant warping.

In either case, the ratio of the peak burst flux to the flux at
touchdown is expected to increase with increasing radius of the
photosphere, as indicated by the observational correlation shown
in Figure~3.

If the origin of the large $f$-ratios is related to interactions of the
burst flux with the disk in the high-inclination sources, we might also
expect a dependence on the orbital phase.
Most of the dippers have only 1--3 radius-expansion bursts each; however,
MXB~1659$-$298 and 4U~1916$-$053 each have 12. We
folded the burst times on the dip/eclipse ephemerides for these two
sources (\citealt{oo01b}, and \citealt{chou00}, respectively), and tested
for variations in the $f$-ratio as a function of orbital phase. In neither
case did we detect significant variation. 
In MXB~1659$-$258 the
three largest $f$-values were all found in the phase bin 0.8--1.0 (i.e.
immediately prior to the eclipse); however, the variation was not
significant compared to the scatter on the $f$-values in the other phase
bins.

\section{Discussion}
\label{conc}

In this paper, we studied \nburst\ photospheric radius expansion bursts
from a large number of accreting neutron stars observed with RXTE to
determine the systematic uncertainties that affect the measurement of
their masses, radii, and distances using the touchdown (Eddington)
fluxes. Comparing the peak fluxes to the touchdown fluxes of the
bursts in our sample, we found that for the majority of the neutron
stars, this ratio is $f \lesssim 1.6$, which is expected on the
grounds of the changing gravitational redshift during the radius
expansion episode \cite[]{pa86}.
However, for some
neutron stars, this ratio can be as large as $f \approx 7$.  We found
that the only sources for which the ratio of these two fluxes is
larger than $f \gtrsim 1.6$ are dippers and Cyg X-2.

For the dippers, we attribute the large $f$-ratios to geometric effects
related to their high inclinations. 
Determining which of the two candidate mechanisms gives rise to the large
ratios may be possible via more detailed X-ray spectroscopic analysis.
For example, time-resolved spectral analysis of the
radius-expansion bursts from dipping sources may provide evidence of a
reflection component from the disk, although the necessity of deconvolving
the changing contribution of this component from the underlying burst
spectral evolution likely presents a considerable challenge.
Such additional analyses are beyond the scope of this paper.

Cyg~X-2 is the only source exhibiting bursts with large $f$-ratios
($>1.6$) that does not also exhibit dips. This source, however, is
distinct from the dip sources in several other respects. 
Cyg~X-2 is the prototypical Z-source 
\cite[]{hvdk89}
and is thought to accrete at a near-Eddington rate, as inferred from
observations of its accretion flow in the UV 
\cite[]{vrtilek90}. This is in sharp contrast to the dippers, all of which
accrete at subtantially sub-Eddington rates 
\cite[e.g.][]{lmxb07}.
The high accretion rate of this source may influence its behaviour in two
different ways.

First, we consider the possibility that the X-ray bursts
observed from Cyg~X-2 are of a different nature than those of the
other bursting sources 
\cite[see][for models of bursts in near-Eddington sources]{twl96}. Indeed,
the bursts in the two high accretion rate sources, Cyg~X-2 and GX~17$+$2,
do not show, at first glance, the characteristic patterns of spectral
evolution seen in the other bursters, making their
classification as type~I or type~II bursts difficult 
\cite[]{szt86,kuul95}. This discrepancy has
been attributed, however, to artifacts introduced by the subtraction
of the accretion flux, which is comparable to the burst
emission in high-luminosity sources.

Second, the large difference in the inferred accretion rates suggest that the
accretion disk may be significantly thicker in Cyg~X-2 than in the dip
sources.  As a result, the geometric effect discussed in \S3.2 for the
dippers may also be responsible for the large $f$-ratios found in Cyg~X-2,
if the latter is observed at a reasonably high inclination.

Modeling the optical lightcurve of the binary constrains the
inclination of Cyg~X-2 to $50^\circ<i<75^\circ$
\cite[]{cow79,ok99}. Such large values of the inclination are also
strongly supported by the secular variations in the X-ray flux of the
source
\cite[]{kuul96} as well as by the super-orbital
periodicities found in its long-term lightcurve 
\cite[]{wij96c}. The combination of a relatively high inclination with
a geometrically thick disk perhaps makes Cyg~X-2 similar to the dippers for
the purposes of the observational effect reported here.

The discrepancy between the maximum and touchdown fluxes for the
high-inclination systems has implications for the measurement of neutron
star masses and radii for bursting neutron stars.
For example, the method recently proposed by \cite{ozel06} 
relies on the accurate measurement
of the Eddington limit at touchdown. Because of this, it is, in
principle, affected by systematic effects on the touchdown flux as the
one presented here. However, the inferred values of both the mass and
the radius of the neutron star depend only on the ratio $F_{\rm
cool}/F_t$ of the flux during the cooling tail of the burst to
that during touchdown. Independent of the particular interpretation of
the unusually large $f$-ratios observed in dippers, this ratio should
remain unchanged because the bias depends on the size of the
photosphere and hence affects in the same way both the cooling and the
touchdown fluxes.  The correlation presented in Figure~\ref{fvsr} further
supports this conclusion.

In contrast, we note that the distance $D$ depends upon the ratio of the
square root of the cooling flux to the Eddington flux, so that
$D$ is affected by the precise value of the Eddington flux.
According to our first geometrical interpretation (where the touchdown
flux is lower than the true Eddington flux due to geometric screening)
the distance derived from the touchdown flux will be too large by a factor
$f^{1/2}$.  
For
EXO~0748$-$676, this would indicate that the distance derived by
\cite{ozel06} should be scaled by
$\sqrt{\left<f\right>}$ to yield a revised
distance of $7.1\pm1.2$~kpc, using the mean ratio $\left<f\right>=1.7\pm0.4$ we
calculate for this source. 
On the other hand, in the second geometric interpretation, the
touchdown flux represents the true Eddington flux and the distance
derived using this value is not affected.  Thus, taking into account
both possible origins for the discrepancy between the peak and
touchdown fluxes, the distance to EXO~0748$-$676 is in the range
6--10~kpc.

It is also worth considering other possible sources of error on
the mass, radius and distance derived via the touchdown flux and
burst tail emission.  In a previous study which attempted to
determine the gravitational redshift from the variation of the burst
flux during photospheric radius-expansion episodes, \cite{damen90}
concluded that systematic effects arising from variations in the
{\it (i)} persistent emission, {\it (ii)} the shape of the burst spectrum, 
and {\it (iii)} photospheric composition could not be neglected. 
The geometric effect described here illustrates that these three additional
factors will only influence the derived mass and radius if they alter the
ratio of the cooling flux to the Eddington flux, and the distance if
they alter the ratio of the square root of the cooling flux to the
Eddington flux.

A possible change in the persistent emission during the burst may
indeed affect 
either the touchdown flux or the cooling flux.
Given that this effect will be most significant for
sources with high persistent luminosities such as Cyg X-2 (as also
shown by Damen et al.  1990), using low persistent luminosity sources
to determine the neutron star mass and radius minimizes this
uncertainty.

As for the second effect, assuming the observed spectra are consistent
with a blackbody (which is usually the case) the 
spectral shape will affect the derived fluxes only if
the spectrum deviates significantly from a blackbody outside the
instrumental bandpass.
In the case
of bursts observed with RXTE, the contribution of flux from above and
below the bandpass is at most 10\%, which likely represents a
conservative limit on the degree of this effect.

Finally, theoretical burst models as well as observations of 4U 1636-536 
suggest that variations in composition may occur in the beginning of 
the burst, prior to the peak. 
In the touchdown and
cooling phases, such variations are not expected. Moreover, the color
correction factors during this cooling stage, which enter the
calculation of neutron star mass and radius, depend very weakly on
metallicity (\"Ozel 2006), 
thus limiting any additional bias.

\section*{Acknowledgments}

F.O. acknowledges support from NSF grant AST-0708640.


\bsp

\label{lastpage}

\end{document}